\documentclass[lettersize,journal]{IEEEtran}
\IEEEoverridecommandlockouts

\usepackage{amsmath,amsfonts}
\usepackage{array}
\usepackage[caption=false,font=normalsize,labelfont=sf,textfont=sf]{subfig}
\usepackage{textcomp}
\usepackage{stfloats}
\usepackage{url}
\usepackage{verbatim}
\usepackage{graphicx}
\usepackage{cite}
\usepackage{xcolor}
\usepackage{booktabs}
\usepackage{hyperref}
\usepackage{url}
\usepackage{here}
\usepackage{algorithm}
\usepackage{amssymb} 
\usepackage[noend]{algpseudocode}
\algrenewcommand\algorithmicrequire{\textbf{Initiate}}
\usepackage{tabularx}
\usepackage[bottom]{footmisc}
\usepackage{comment}
\usepackage{nomencl}
\newcommand{\etal}{\emph{et~al.\vspace{-0.25px}}\@}
\usepackage{adjustbox}
\usepackage{subcaption}

\definecolor{darkgreen}{RGB}{0,100,0}

\def\BibTeX{{\rm B\kern-.05em{\sc i\kern-.025em b}\kern-.08em
    T\kern-.1667em\lower.7ex\hbox{E}\kern-.125emX}}

\bstctlcite{bstctl:nodash}

\makenomenclature

\begin{document}

\title{Active Learning of Fractional-Order Viscoelastic Model Parameters for Realistic Haptic Rendering}

\author{Harun Tolasa${}^*$,~\IEEEmembership{Student Member,~IEEE}
\hspace{25mm}  Gorkem Gemalmaz${}^*$,~\IEEEmembership{Student Member,~IEEE} \\ \vspace{3mm}
Volkan Patoglu,~\IEEEmembership{Member,~IEEE}
\thanks{H. Tolasa, G. Gemalmaz, and V. Patoglu are with the Faculty of Engineering and Natural Sciences at Sabanci University,  Istanbul, Turkiye.\\ ${}^*$~The first two authors have contributed equally to this study.\\ {\tt\scriptsize \{harun.tolasa,gorkem.gemalmaz,volkan.patoglu\}@sabanciuniv.edu}}}

\markboth{Transactions on Haptics,~Vol.~xx, No.~xx, xx~xx}
{Shell \MakeLowercase{\textit{et al.}}: A Sample Article Using IEEEtran.cls for IEEE Journals}

\maketitle

\begin{abstract} 
Effective medical simulators necessitate realistic haptic rendering of biological tissues that exhibit viscoelastic material properties, such as creep and stress relaxation. Fractional-order models provide an effective means of describing intrinsically time-dependent viscoelastic dynamics with few parameters, as they naturally capture memory effects. However, due to the unintuitive, frequency-dependent coupling among the order of the fractional element and other parameters, determining appropriate parameter values for fractional-order models that yield high perceived realism remains a significant challenge.  In this study, we propose a systematic means of determining the parameters of fractional-order viscoelastic models that optimizes the perceived realism of haptic rendering across general populations.  First, we demonstrate that the parameters of fractional-order models can be effectively optimized through active learning, using qualitative feedback-based human-in-the-loop~(HiL) optimization, to ensure consistently high realism ratings for each individual. Second, we propose a rigorous method to combine HiL optimization results into an aggregate perceptual map trained on the entire dataset, and demonstrate how to select population-level optimal parameters from this representation that are broadly perceived as realistic across general populations. Finally, we provide evidence of the effectiveness of the generalized fractional-order viscoelastic model parameters for three viscoelastic materials by characterizing their perceived realism through human-subject experiments. Overall, generalized fractional-order viscoelastic models established through the proposed HiL optimization and aggregation approach possess the potential to significantly improve the sim-to-real transition performance of medical training simulators.
\end{abstract} \vspace{-1mm}


\vspace{-1mm}
\section{Introduction}

Haptic rendering is the process through which virtual objects are made tangible to users via a haptic interface that provides kinesthetic feedback. Haptic rendering has widespread applications in driver assistance systems, computer games, and simulators used for motor skill training, including medical simulators~\cite{salisbury1995haptic}. Medical simulators enable trainee surgeons to practice complex procedures in virtual environments without exposing health risks. For instance, haptic rendering is employed to train laparoscopic procedures.  
Haptic rendering also finds applications in medical diagnostics training, where identifying abnormalities in tissue mechanics caused by diseases such as cancer is crucial~\cite{breast_cancer,breast_tissue}. Medical simulators that accurately render viscoelastic tissue behavior can help train physicians more effectively by improving the performance of the sim-to-real transition.

Biological tissues possess viscoelastic material properties that are intrinsically time-dependent~\cite{time_dependant_tissue}. This behavior makes their modeling challenging. Traditional integer-order spring-damper networks fail to provide satisfactory approximations, as they cannot capture the memory-dependent nature of viscoelasticity~\cite{IO_fails_memory}. Consequently, \emph{linear} fractional-order models are widely embraced as more accurate representations. Fractional-order models can describe sophisticated viscoelastic dynamics with fewer parameters, naturally capturing long-time memory effects~\cite{FO_few_param}. On the other hand, fractional-order models are significantly less intuitive; while physical interpretations of fractional-order models are available, the effects of model parameters on material behavior are harder to predict due to the frequency-dependent coupling between the fractional order and the other parameters. Consequently, determining appropriate viscoelastic model parameters to achieve high perceived realism in haptic rendering remains a significant challenge, limiting the widespread adoption of such models.

In this study, we first provide a systematic approach to determine the parameters of fractional-order viscoelastic models to optimize their perceived realism for each individual. While custom parameters for individuals can help achieve personalized experiences with high perceived realism, such results do not necessarily generalize well to other users, as each individual differs in perception, sensitivity, and interaction style. To determine model parameters that are broadly perceived as realistic across the general population, it is necessary to capture the underlying perceptual mapping from a diverse group of individuals. Such generalized characterizations can enable researchers to uncover consistent perceptual preferences within a population, supporting the development of haptic systems that perform effectively over a diverse set of users, rather than being fine-tuned for specific individuals. 

Accordingly, we extend our approach to construct an aggregate perceptual map trained on the entire dataset that captures the perceptual preferences of the population for any feasible parameter set. Once the aggregate perceptual map is determined, we use this map to select the model parameters that maximize perceived realism at the population level. Consequently, our approach not only provides realistic estimates of fractional-order viscoelastic model parameters for specific individuals but also proposes a rigorous method to form an aggregate perceptual map and to estimate population-level optimal parameters that are perceived as realistic across general populations. 

\smallskip
Our contributions can be summarized as follows: 
\smallskip
\begin{itemize}

\item[(i)] We propose an active learning framework, based on sample-efficient human-in-the-loop~(HiL) Bayesian optimization with qualitative perceptual feedback, where users' feedback is collected iteratively to determine the parameters that maximize the perceived realism of viscoelasticity renderings.

\item[(ii)] Given the perceptual mappings characterized for each individual, we propose a rigorous means to combine these results so that an aggregate perceptual map trained on the entire dataset is computed. Once individual preferences are mapped into an interpretable population-level representation, we select population-level optimal parameters for this representation that are perceived as realistic. 

\item[(iii)] We provide evidence of the effectiveness of the generalized fractional-order viscoelastic model by characterizing its perceived realism through human-subject experiments. 
\end{itemize}

\smallskip
Overall, fractional-order viscoelastic models established through the proposed HiL optimization and aggregation approach possess the potential to substantially enhance the realism of viscoelasticity renderings, such as tissue rendering in medical training simulators.

\section{Related Work} \label{Related_work}

\smallskip
\subsection{Viscoelastic Models}

Viscoelastic materials exhibit intricate time-dependent phenomena, most notably creep and stress relaxation. Creep is ongoing deformation under fixed stress, while stress relaxation is the disappearance of stress over time under fixed strain~\cite{Ferry1980}. Classical integer-order models, i.e., the Maxwell and Kelvin–Voigt models, can partially describe one of these phenomena but cannot describe both simultaneously with reasonable accuracy. To alleviate this shortfall, Zener proposed the Standard Linear Solid~(SLS) model~\cite{Zener1948}, which combines components of the Maxwell and Kelvin–Voigt models. The SLS model improves the predictive capability for both creep and stress relaxation, but still cannot model materials whose time responses do not follow simple exponential forms.

One crucial development in viscoelastic modeling was achieved through the use of fractional calculus. Early attempts by Gemant~\cite{Gemant1936} recognized that fractional derivatives can more naturally describe hereditary effects in material response. This idea was later extended and developed by Bagley and Torvik~\cite{BagleyTorvik1979, BagleyTorvik1983}, who demonstrated that fractional-order models can effectively capture creep and stress relaxation with fewer parameters and greater consistency than more complex integer-order models. These developments gave rise to the fractional-order SLS, in which the viscous damper is substituted by a fractional spring-damper element of order $\alpha$, where $0 \!< \! \alpha \!<\! 1$.  This extension enables power-law-type time responses that better capture the intrinsic memory-dependent behavior of viscoelastic materials.

Fractional-order viscoelastic models have found widespread applications in biological and biomedical engineering. Human soft tissues, such as muscle and fat, routinely exhibit viscoelastic behavior that is poorly characterized by integer-order models, but well-approximated by fractional-order tissue models~\cite{Koeller1984}. The high fidelity of fractional-order models, which require only a few parameters, has made them a preferred choice for modeling soft tissues, where time-dependent memory effects are fundamental to the material response.

The importance of realistic tissue mechanics for haptic medical simulators has been highlighted in~\cite{BasdoganSrinivasan2002}. Fractional-order models have also been employed for haptic rendering~\cite{tokatli2015stability, TokatliPatoglu2015} and interaction control during physical human–robot interaction~\cite{Aydin2017, Aydin2018, Aydin2020, Sirintuna2020}. In particular, fractional-order models were proposed for haptic rendering, and the coupled stability and/or passivity of rendering such models was studied in~\cite{tokatli2015stability, TokatliPatoglu2015}. Recently, authors have extended these studies to provide analytical passivity bounds for haptic rendering with a fractional-order SLS model under short memory discretization, which generalizes previously reported results on passive rendering of viscoelasticity~\cite{Gemalmaz2026}. Similarly, fractional-order admittance controllers were proposed, and the complementary stability of this interaction control strategy was studied in~\cite{Aydin2017, Aydin2018, Aydin2020, Sirintuna2020}. Overall, these studies provide evidence that fractional models possess the potential to help bridge the gap between viscoelastic tissue response and its high-fidelity haptic renderings.

\subsection{HiL Bayesian Optimization in Haptic Studies}

Bayesian optimization with Gaussian processes~(GPs) has been extended beyond regression tasks based on numerical measurements to active learning frameworks that rely on qualitative feedback~\cite{williams1998, Rasmussen2005, Chu2005}. Early work on qualitative feedback applied GPs to binary decisions, such as ``yes/no'' responses~\cite{williams1998}. This approach was later generalized to handle ordinal labels and pairwise preference feedback~\cite{Chu2005}. These extensions enabled the application of GP-based Bayesian optimization in diverse domains, including psychophysical experiments~\cite{Owen2021, Gardner2015, Schlittenlacher2018, Song2018, Browder2019} and HiL optimization studies~\cite{Tucker2020, Biyik2020, Li2020, Tucker2020_2, Tolasa2026}.  

Compared to a broader set of applications, research on HiL optimization using qualitative perceptual feedback in haptics has been relatively limited. Notable examples include HiL optimization for haptic rendering~\cite{Catkin2023}, texture synthesis~\cite{Lu2022}, visual-haptic parameter matching~\cite{Ujitoko2023} and multi-modal perception under sensory conflicts~\cite{Tolasa2024}. In particular, Catkin~\etal~\cite{Catkin2023} introduced a preference-based HiL Bayesian optimization method to enhance the perceived realism of spring and friction rendering, while Tolasa~\etal~\cite{Tolasa2024} optimized parameters affecting visual-haptic congruency using qualitative perceptual feedback.

In addition to HiL Bayesian optimization methods, alternative approaches have been proposed for tuning haptic interfaces, including interactive recommendation systems~\cite{Theivendran2024}, and generative model-based learning methods~\cite{Zhang_Mingxin2025}.

The present study is in the same spirit as~\cite {Catkin2023,Tolasa2024}, as it focuses on improving the perceived realism of haptic rendering using GP-based HiL optimization methods. On the other hand, this study is significantly different as it focuses on viscoelastic model rendering based on non-intuitive fractional-order models, and proposes a rigorous method to train an aggregate perceptual map across the entire dataset of various users and to determine population-level model parameters that are broadly perceived as realistic.

\subsection{Aggregation Methods}\smallskip

Statistical averaging of the outcomes of individual GP models can provide a summary of participants’ perceptions~\cite{Tolasa2024}. However, averaging assigns equal weight to all individual GP models, without any consideration of whether an individual GP model is constructed over a well-sampled or sparsely-sampled region. This uniform weighting can undesirably pull the averaged latent scores toward their priors and artificially vanish the averaged variance, even in unsampled regions. Consequently, the averaged model’s ability to generalize the relationship between the parameters and latent scores may deteriorate with statistical averaging. 

On the other hand, fitting a single GP model to the aggregated dataset can capture generalizable patterns across participants, as this model is trained on the entire dataset~\cite{Rudovic2019,tanaka2019}. This approach preserves uncertainty in sparsely sampled regions while maintaining consistent latent scores across the parameter space. Hence, the aggregate GP model’s ability to generalize the relationship between the parameters and latent scores is preserved~\cite{Tresp2000}.

However, training a single GP model on the pooled dataset is neither computationally efficient nor practical, as a prominent drawback of GP models is their limited scalability~\cite{Rasmussen2005,Tresp2000,Snelson2005}. To handle training with large datasets, a variety of scalable approximation strategies have been proposed. These strategies can be loosely categorized into three groups~\cite{Liu2020}: sparse approximations, low-rank matrix factorizations, and ensemble prediction methods. Sparse methods reduce computational cost by modeling the data using a small representative subset~\cite{Snelson2005,QuinoneroCandela2007,Titsias2009}. Low-rank approaches truncate the kernel matrix to exploit linear algebraic shortcuts~\cite{Nguyen2019a,Csato2002}. Ensemble prediction methods, such as the Bayesian Committee Machine~(BCM), partition the data set into smaller subsets and train different GP models on these subsets, so that each GP model provides a local posterior. Finally, these GP models are combined to approximate the estimate of a global posterior~\cite{Tresp2000}.

In this study, we use BCM to develop an aggregate posterior GP model, as this method is well-suited to handling data from human-subject experiments. In such experiments, each participant contributes an independent, (almost) equal-sized subset of the complete dataset, acquired under identical experimental conditions. BCM can leverage this already-partitioned data structure by training a distinct GP model for each participant and then merging their posteriors into a single prediction using Bayes' rule, assuming conditional independence. 

Consequently, BCM is particularly well-suited to aggregate data from human-subject experiments, as it respects each participant’s unique preferences while capturing patterns shared across the group. In doing so, it also provides a data-efficient means of mapping individual preferences into an interpretable population-level representation, also called the \emph{aggregate perceptual map}. Once the aggregate perceptual map is constructed, this map can be used to select the optimal parameter sets for a given perception task.

\newpage
\section{Fractional-Order Model of Viscoelasticity}

\begin{figure}[h!]
    \centering 
    \includegraphics[width=.5\linewidth, trim=10 17 10 20, clip]{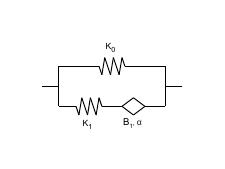}
    \vspace{-2mm}
    \caption{Fractional-order standard linear solid model}
    \label{fig:FO-SLS}
\end{figure}

The fractional-order SLS model, depicted in Figure~\ref{fig:FO-SLS}, generalizes the well-known SLS model with a fractional-order derivative, whose order is between 0 and 1. 

There exist several different (and equally valid) definitions for fractional derivatives. In this study, we use the Grünwald–Letnikov derivative~\cite{grunwald} because it can be easily adapted to discrete-time realizations. The Grünwald–Letnikov derivative is defined as \vspace{-1mm}

\begin{equation}
D^\alpha f(x) = \lim_{h \to 0} \frac{1}{h^\alpha} \sum_{k=0}^\infty (-1)^k \binom{\alpha}{k} f\big(x - kh\big) \nonumber
\end{equation}

\noindent where $\alpha$ is the order of the derivative, $h$ is the step size, and $\binom{\alpha}{k}$ denotes the generalized binomial coefficient, with 

\begin{equation}
\binom{\alpha}{k} = \frac{\Gamma(\alpha+1)}{\Gamma(k+1)\Gamma(\alpha-k+1)} \nonumber
\end{equation}

\noindent while $\Gamma$ represents the Gamma function.

Fractional-order derivatives include standard (integer-order) derivatives as special cases; e.g., $D^1$ denotes the first derivative, and $D^0$ is the identity operator. 

Unlike integer-order derivatives, fractional-order derivatives are \emph{non-local}; fractional derivatives depend on the entire history of the function. Accordingly, systems described by fractional derivatives can exhibit memory-dependent dynamics, such as creep and stress relaxation in viscoelastic materials.
    
Furthermore, fractional-order derivatives are \emph{linear} operators with well-defined Laplace transforms. If $D^{\alpha}f(t)$ represents the Grünwald-Letnikov fractional derivative of order $\alpha$, then its Laplace transform is given by \vspace{-1mm}

\begin{equation}
\mathcal{L}\{D^{\alpha}f(t)\} = s^{\alpha}F(s) - \sum_{k=0}^{n-1}s^{\alpha-1-k}f^{(k)}(0^+) \nonumber
\end{equation}

\noindent where $n = \lceil \alpha \rceil$ denotes the smallest integer greater than or equal to $\alpha$. Assuming that the initial conditions $f^{(0)}(0^+)$ and all its derivatives up to order $\lceil \alpha \rceil - 1$ are zero (as in the computation of transfer functions), the Laplace transform simplifies to \vspace{-2mm}

\begin{equation}
\mathcal{L}\{D^{\alpha}f(t)\} = s^{\alpha} \, F(s)  \nonumber
\end{equation} 

\noindent which is an important generalization in the Laplace domain.

The force-position impedance transfer function~$H(s)$ of the fractional-order~SLS model in Figure~\ref{fig:FO-SLS} can be computed as

\begin{equation}
H(s) = K_0 + \frac{K_1 B_1 s^\alpha}{K_1 + B_1 s^\alpha} \nonumber
\end{equation}

\noindent where $K_0$ and $K_1$ denote stiffness constants, $\alpha$ represents the order of the fractional-order derivative, and $B_1$ is the coefficient of the fractional-order element, which simultaneously displays stiffness and damping properties for $\alpha \in (0,1)$. In this study, the proposed HiL optimization framework was used to determine $K_1$, $B_1$, and $\alpha$, while $K_0$ was determined using the effective-stiffness constraint imposed on the rendering.

\medskip

\begin{figure*}[t!]
    \centering    
    \includegraphics[width=.95\textwidth]{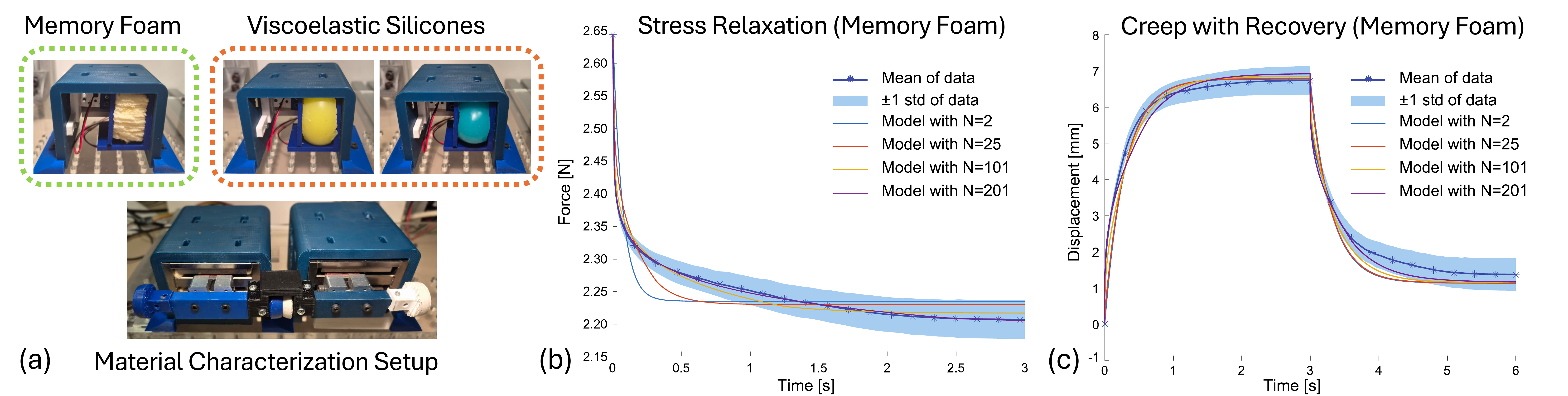}
    \vspace{-1.mm}
    \caption{(a)~Tested viscoelastic materials and the material characterization setup. Experimental data for the memory foam material for (b)~the stress relaxation experiments, and (c)~the creep with recovery experiments.  Best fits of the fractional-order SLS model for different memory lengths are also shown. Characterization results for the other two viscoelastic silicone materials are provided in the Supplementary Document~\cite{Supplementary}.} \vspace{-2mm}
    \label{fig:systemID}
\end{figure*}

\subsubsection*{Short-Memory Discretization}

A practical concern arises when implementing fractional-order derivatives: while formal definitions require an infinite history, the number of memory terms~$N$ used in practical computations cannot be infinite. The series must therefore be truncated at a finite memory length. The short-memory discretization method~\cite{short_memory} provides a solution to this problem, based on the observation that the values of the binomial coefficients in the Grünwald-Letnikov definition are largest around the most recent data and decay rapidly for past observations. Accordingly, the short-memory discretization method considers the most recent data inside a finite window to approximate the fractional-order derivative.

According to short-memory discretization, the z-transform of the fractional-order derivative of order $\alpha$ for $x(t)$ is given by  \vspace{-3mm}

\begin{equation}
\mathcal{Z}\left\{D^{\alpha} [x(t)]\right\} \approx \frac{1}{T^{\alpha}} \sum_{i=0}^{N} c_i \, z^{-i} \nonumber
\end{equation}

\noindent where $T$ is the sampling time, $N$ is the window length, and the coefficients $c_i$ can be computed recursively according to \vspace{-3mm}

\begin{align*}
c_{0} &= 1, \\
c_{i} &= (-1)^{i} \binom{\alpha}{i} = \frac{i-\alpha-1}{i} \cdot c_{i-1}, \quad \text{for } i = 1, 2, \ldots, N. 
\end{align*} \normalsize

When $\alpha = 1$, this formula degenerates into the well-established backward-difference discretization method.

Consequently, the discrete force–position impedance transfer function for the fractional-order SLS virtual environment model can be expressed as \vspace{-2mm}

\begin{equation}
H(z) = K_0 + \frac{K_1 B_1 \frac{1}{T^{a}} \sum_{i=0}^{N} c_i z^{-i}}{K_1 + B_1  \frac{1}{T^{a}} \sum_{i=0}^{N} c_i z^{-i}} \nonumber
\end{equation}

In this discrete model, the choice of window length~$N$ is significant, as it affects how well the model captures stress relaxation and creep response.

\section{Characterization of the Reference Materials}

During the experimental evaluation of perceived realism, detailed in Section~\ref{Experiments}, physical viscoelastic materials, as shown in Figure~\ref{fig:systemID}(a), served as the physical references against which all comparisons were made. System identification was conducted on these materials to characterize baseline parameter values for their fractional-order models through “creep with recovery” and “stress relaxation” tests. For this purpose, two identical direct-drive haptic interfaces were mechanically coupled via a rigid connection, as in Figure~\ref{fig:systemID}(a), where one device applied controlled force/motion inputs while the resulting displacements/interaction forces were measured.

For brevity, only the results for the memory foam material are presented in the manuscript, while the results for the other two viscoelastic silicone materials with medium and high stiffness properties are presented in the Supplementary Document~\cite{Supplementary}, to demonstrate the applicability of the proposed framework across different material properties. In Section~\ref{Experiments}, the identified parameters of all three viscoelastic materials were used as baselines for comparison with the parameters obtained via the active learning and aggregation framework in the human subject study.

In the creep with recovery tests, a constant force of 3~N was applied to the material, and the resulting deformation was observed for 3~s. Then, the force was suddenly reduced to 0.5~N, and the resulting deformation was observed during the recovery phase.  In stress relaxation tests, the deformation was set to 5~mm, and the reaction force of the material was monitored for 3~s. Each test was repeated 16~times. Figure~\ref{fig:systemID} presents the mean and standard deviation of the data collected during these experiments, as well as several best fits of the fractional-order SLS~model.

In the discrete-time implementation of the fractional-order SLS~model, the choice of window length~$N$ significantly affects how well the model captures the stress relaxation and creep response of the viscoelastic material.  In order to determine an appropriate value of $N$, best fits were obtained for various window lengths $N \in \{2, 25, 101, 201\}$. These model fits are presented in Figure~\ref{fig:systemID}. The quality of the model fits was quantified using normalized root-mean-square error~(NRMSE).  

As evident in Figure~\ref{fig:systemID}, the quality of the fit of the model increased significantly with higher values of~$N$.  At $N = 101$, reasonable accuracy was achieved with less than $4\%$ NRMSE:  The error was $3.92\%$ for creep with recovery and $0.32\%$ for stress relaxation.  Consequently, a window size of 101 was selected for subsequent analyses and experiments, as it achieved a sufficiently low modeling error while remaining more computationally efficient than larger window lengths, which offered only marginal improvements in accuracy. 

The corresponding model parameters for the memory foam were determined as: $K_0~=~-2.89$~N/mm, $K_1 = 5.70$~N/mm, $B_1~=~5.89$~N$\cdot$s$^{\alpha}$/mm, and $\alpha = 0.203$. Here, $K_0$ is negative, as this stiffness compensates for the extra stiffness introduced by the fractional-order element. The identified model parameters result in a passive model, indicating a realistic representation of the physical viscoelastic material. The passivity of the sampled-data system with the discrete-time FO-SLS model under short-memory discretization is validated both numerically and analytically, according to the bounds established in~\cite{Gemalmaz2026}.

\section{Human-in-the-Loop Bayesian Optimization with Qualitative Perceptual Feedback}
\smallskip
\subsubsection{Gaussian Process Model for Perceived Realism}

The parameter space of the haptic rendering model is defined as $A=\{x \in \mathbb{R}^d : 0 \le x_i \le 1\}$. Let $f(x)$ denote the latent function of perceived realism. We place a Gaussian process prior on $f(x)$ as \vspace{-2mm}

\begin{equation}\label{prior f PDF Eq}
f(\boldsymbol{x}) \sim \mathcal{GP}(0,K), \nonumber
\end{equation} 

\noindent where $K \in \mathbb{R}^{n\times n},\, k_{i,j}=k(x_i,x_j)$ is the noiseless kernel matrix of the GP regression model.

Let $\boldsymbol{q}=\{q_{1},q_{2},...,q_{n}\}$ be the participant's ordinal classification feedback, and define the dataset as $\boldsymbol{D}=\{(x_i,q_i)\}_{i=1}^n$. The posterior is obtained from
\begin{equation}
P(f|\boldsymbol{D}) \propto P(\boldsymbol{D}|f)P(f),  \nonumber
\end{equation}
where $P(f)$ is the GP prior and $P(\boldsymbol{D}|f)$ is the likelihood of the feedback.

Following~\cite{Li2020,Tolasa2024}, ordinal classification feedback was modeled as $O=\{o_1,o_2,o_3\}$, corresponding to ``different'', ``similar'', and ``close''. Let $t=\{t_0,t_1,t_2,t_3\}$ be ordered thresholds with $-\infty=t_0 < t_1 < t_2 < t_3=\infty$. Then, the probability of classifying $x_i$ as $o_j$ is \vspace{-2mm}

{\small
\begin{equation}\label{Ordinal Class Prob Eq}
P(q_i=o_j|f)=\Phi\!\left(\frac{t_j-f(x_i)}{c_o}\right)-\Phi\!\left(\frac{t_{j-1}-f(x_i)}{c_o}\right)  \nonumber
\end{equation}}
\normalsize

\noindent where $\Phi$ is the Gaussian cumulative distribution function and $c_o>0$ models classification noise. Assuming independence of feedback, one can state that \vspace{-2mm}

\begin{equation}\label{eqn: Qualitative Dataset Probability}
P(\boldsymbol{D}|f)=\prod_{i=1}^{n}P(q_i|f).  \nonumber
\end{equation}

The posterior was approximated as a multivariate Gaussian distribution using the Laplace method~\cite{Rasmussen2005} to enable Bayesian optimization with qualitative human feedback, as detailed in~\cite{ Biyik2020, Tucker2020, Li2020, Tolasa2024}. 

For any arbitrary parameter $x_*$ in the search space, the posterior mean and variance of the corresponding latent score $f_*$ can be calculated as

\noindent
\begin{eqnarray}
\mu_{*|\boldsymbol{D}} &=& k_{*,1:n} \, K^{-1}\hat{f}  \label{Inference Expectation} \nonumber \\
\sigma_{*|\boldsymbol{D}}^2 &=& k_{**}-k_{*,1:n} \, (K+W^{-1})^{-1} k_{*,1:n}^{T}. \label{Inference Variation} \nonumber
\end{eqnarray}

The $\hat{f}$ is the latent score that maximizes the probability of the given qualitative feedback
\begin{equation}\label{Posterior Expectation}
\hat{f}= \text{argmax}_{f}\left(log(P(\boldsymbol{D}|f)P(f))\right) \nonumber
\end{equation}
\noindent and $W$ is the negative Hessian of $log(P(q|\hat{f}))$ defined as
\begin{equation}\label{Negative Hessian}
W_{ij}=-\frac{\partial^2 \, log(P(D|\hat{f})}{\partial\hat{f_i} \, \partial \hat{f_j}}. \nonumber
\end{equation}

\smallskip
\subsubsection{HiL Bayesian Optimization}

Algorithm~\ref{alg:ALBO} summarizes HiL Bayesian optimization. The first $N_s$ trials explored the parameter space with a space-filling design. The remaining $N\!-\!N_s$ iterations selected parameters via the acquisition function. Each trial provided outcome data used to update the GP posterior. At the end, the algorithm outputs $x_{\max}$ with the highest posterior mean.

\algrenewcommand\algorithmicrequire{\textbf{initiate}} 
\begin{algorithm}[h!] 
\caption{HiL Bayesian Optimization}
\label{alg:ALBO}
\begin{algorithmic}[1]
\Require {$S$: Parameter space, $f\sim GP(\mu_0, \sigma_0)$: GP prior, $N_s$: Space-filling iterations, $N$: Total iterations}
\For{$i=1,2,\dotsc,N$}
    \If{$i \leq N_s$}
        \State Select $x_i$ from $S$ using space-filling
    \Else
        \State Select $x_i$ that maximizes the acquisition function
    \EndIf
    \State Observe outcome $d_i$ and add to dataset $D$
    \State Update GP posterior using $D$
\EndFor
\State Return $x_{\max}$ where $f(x_{\max})$ has the largest posterior mean
\end{algorithmic}
\end{algorithm} \normalsize

\subsubsection{GP Hyperparameters, Search Space, and Sampling Strategy}
For developing a GP model, a radial basis function, $k_{{i,j}}~=~\exp(-\theta{||x_i-x_j||}_2^2)$, was used as the kernel function with $\theta$ selected as 30. Qualitative perceptual feedback was modeled using an ordinal likelihood with noise parameter $c_o$ selected as $0.5$ and classification thresholds determined as \(t=\{-\infty,-0.5,0.5,\infty\}\), for three ordered categories.

New parameters are selected using an upper confidence bound~(UCB) acquisition function. To sample a new rendering parameter $x_{n}$ for ${n}^{th}$ HiL trial, $\zeta$ was utilized as \vspace{-4.5mm}

 \begin{eqnarray}
  \zeta(x_*)_n=\mu_{*|\boldsymbol{D_{n-1}}}+\lambda \, \sigma_{*|\boldsymbol{D_{n-1}}}  \nonumber
\end{eqnarray}
 
\noindent where $\mu_{*|D_{n-1}}$ and $\sigma_{*|D_{n-1}}$ denote the posterior mean and standard deviation of an arbitrary parameter set $x_*$ given the data up to the $(n\!-\!1)$\textsuperscript{th} iteration. The parameter $\lambda$ is the UCB exploration weight and was set to $0.7$.

A broad range of fractional-order viscoelastic model parameters was used to define the search space, with stiffness constants $K_0 \in [-15.7,\, 0.44]$~N/mm and $K_1 \in [1,\, 32]$~N/mm, damping constant $B_1 \in [0.001,\, 32]$~\text{N$\cdot$s$^{\alpha}$/mm}, and fractional order $\alpha \in [0.01,\, 0.99]$. To ensure coupled stability of interactions, the feasible parameter sets were confined to the ones that result in the passivity of viscoelastic rendering~\cite{Gemalmaz2026}. Furthermore, to reduce the dimension of the search space, the effective stiffness of the fractional-order model was matched to that of the physical viscoelastic material in the low-frequency range, thereby removing one degree of freedom and reducing the search space to three dimensions. The effective damping was kept within a suitable range to allow participants to perceptually compare creep and stress-relaxation behavior during haptic interaction. Accordingly, $K_1$, $B_1$, and $\alpha$ were used as the free parameters, while $K_0$ was determined via the effective-stiffness constraint. During HiL optimization, the search space was normalized to the $[0,1]$ range for GP modeling and further constrained by the sampled-data passivity condition~\cite{colgate1997}, numerically checked for each parameter set.

The acquisition function for HiL optimization and the GP hyperparameters were chosen a priori based on pilot experiments. The kernel hyperparameter, ordinal classification noise level, and UCB exploration weight were selected to produce smooth latent perceptual functions and reliable behavior within a limited trial budget, consistent with related work in haptics~\cite{Catkin2023, Tolasa2024} and active learning studies in~\cite{Tucker2020, Li2020, Biyik2020}.

The pilot sessions further suggested that approximately 25 trials are typically sufficient to converge toward a parameter region that yields a perceptually plausible match to the viscoelastic behavior of the reference material, in line with previous findings in~\cite{Catkin2023, Tolasa2024}. During the HiL optimization sessions, the first five trials were selected using a Latin hypercube sampling design to obtain an initial space-filling, after which the parameters were sampled according to the acquisition function. To avoid overfitting and to ensure consistent aggregation across participants, all hyperparameters were kept fixed throughout the study.

After completing the human-subject experiments, we performed a post-hoc robustness check by comparing the use of empirically selected GP hyperparameters with the marginal-likelihood-optimized values. Our analysis indicates that both settings yield similar optimal rendering parameter sets with NRMSE less than~3\%, and that the aggregate model is robust to hyperparameter variations.

\vspace{-1mm}
\section{Aggregation of Gaussian Process Models}

The GP posterior models obtained from individual experiments are combined to form a population-level aggregate GP model, which is used to understand how participants collectively perceive realism across different haptic rendering parameter sets. The aggregation of individual GPs to an interpretable population-level representation is performed using the Bayesian Committee Machine~(BCM)~\cite{Tresp2000}. 

BCM is performed as follows: Let $\boldsymbol{D_{tot}}=\{\boldsymbol{D_{s1}},\boldsymbol{D_{s2}},\dots,\boldsymbol{D_{sM}}\}$ be the total dataset of $M$ users. The aggregate posterior is obtained from the proportionality as \vspace{-4.5mm}

\begin{equation}\label{eqn: Qualitative Dataset Probability}
P(f|\boldsymbol{D_{tot}})\propto{\frac{\prod_{j=1}^{M}P(f|\boldsymbol{D_{sj}})}{P(f)^{M-1}}}  \nonumber
\end{equation}

\noindent where $P(f)$ is the GP prior and $P(f|\boldsymbol{D_{sj}})$ is the GP posterior of $j^{th}$ user.  For a single latent score, the BCM approximates its posterior mean as 

\vspace{-5mm}
\begin{equation}\label{eqn: Generalized posterior mean}
\mu_{*|\boldsymbol{D_{tot}}}=\sigma_{*|\boldsymbol{D_{tot}}}^2\sum_{j=1}^{M}\sigma_{*|\boldsymbol{D_{sj}}}^{-2} \, \mu_{*|\boldsymbol{D_{sj}}}\nonumber
\end{equation}

\noindent where the posterior variance is calculated from its inverse, given as \vspace{-4mm}

\begin{equation}\label{eqn: Generalized posterior variance}
\sigma_{*|\boldsymbol{D_{tot}}}^{-2}=-(M-1)k_{**}^{-1}+\sum_{j=1}^{M}\sigma_{*|\boldsymbol{D_{sj}}}^{-2}.\nonumber
\end{equation} \vspace{-2mm}

\noindent with $\mu_{*|\boldsymbol{D_{sj}}}$ and $\sigma_{*|\boldsymbol{D_{sj}}}^{2}$ denoting the mean and variance prediction of $j^{th}$ GP posterior model for any arbitrary point~$x_*$. 

Upon aggregation through BCM, each individually trained GP model captures the perceptual response pattern of a single participant. These individual models are then combined to form a single population-level prediction. The combination is based on precision weighting: GP models that make confident (low-variance) predictions have a stronger influence on the final estimate, while models with high uncertainty contribute with minimal influence. This prevents the population-level predictions from being distorted by sparse, noisy, or less informative data from any individual participant.

\section{Human Subject Experiments} \label{Experiments}

\vspace{-2mm}
\subsection{Task}
The main task assigned to the subjects was to compare the perceptual similarity between a \emph{physical} reference viscoelastic material and the corresponding virtual environment presented through a haptic interface. During each trial, subjects were required to interact with two different devices: the reference haptic interface containing the physical viscoelastic material, and the test interface driven by a virtual viscoelastic model.

Participants were asked to pay attention to the overall haptic feel of the interaction-—stiffness, damping, as well as time-dependent creep and stress relaxation behavior-— and to make their judgments solely on the basis of perceptual similarity. There was no performance-based scoring or correctness criterion; the task was purely perceptual and subjective.

\vspace{-2mm}
\subsection{Participants}

For the sessions with the memory-foam material, 24 participants (mean age: 23.3 years) were recruited for the HiL \emph{Viscoelastic Model Optimization} session, and 20~different participants (mean age: 22.6 years) were recruited for the \emph{Model Validation} session. For the sessions with the viscoelastic silicone materials, 16~participants (mean age: 22.8 years) were recruited for the HiL \emph{Viscoelastic Model Optimization} and \emph{Model Validation} sessions, with 8 participants randomly assigned to experiments with each silicone material. Finally, 8~additional participants (mean age: 22.2 years) were recruited for the \emph{Aggregate-Best Validation} session, and these participants completed the aggregate-best validation experiments for all three reference materials.

All participants were right-handed and reported no known sensory-motor deficiencies. None of the participants had extensive prior experience with haptic interfaces or psychophysical experiments. Before the experimental sessions began, all participants provided written informed consent in accordance with the ethical guidelines approved by the Institutional Review Board of Sabanci University (Protocol No: FENS-2025-17).

\vspace{-4mm}
\subsection{Apparatus}

\begin{figure}[t!]
    \centering
    \includegraphics[width=.95\linewidth]{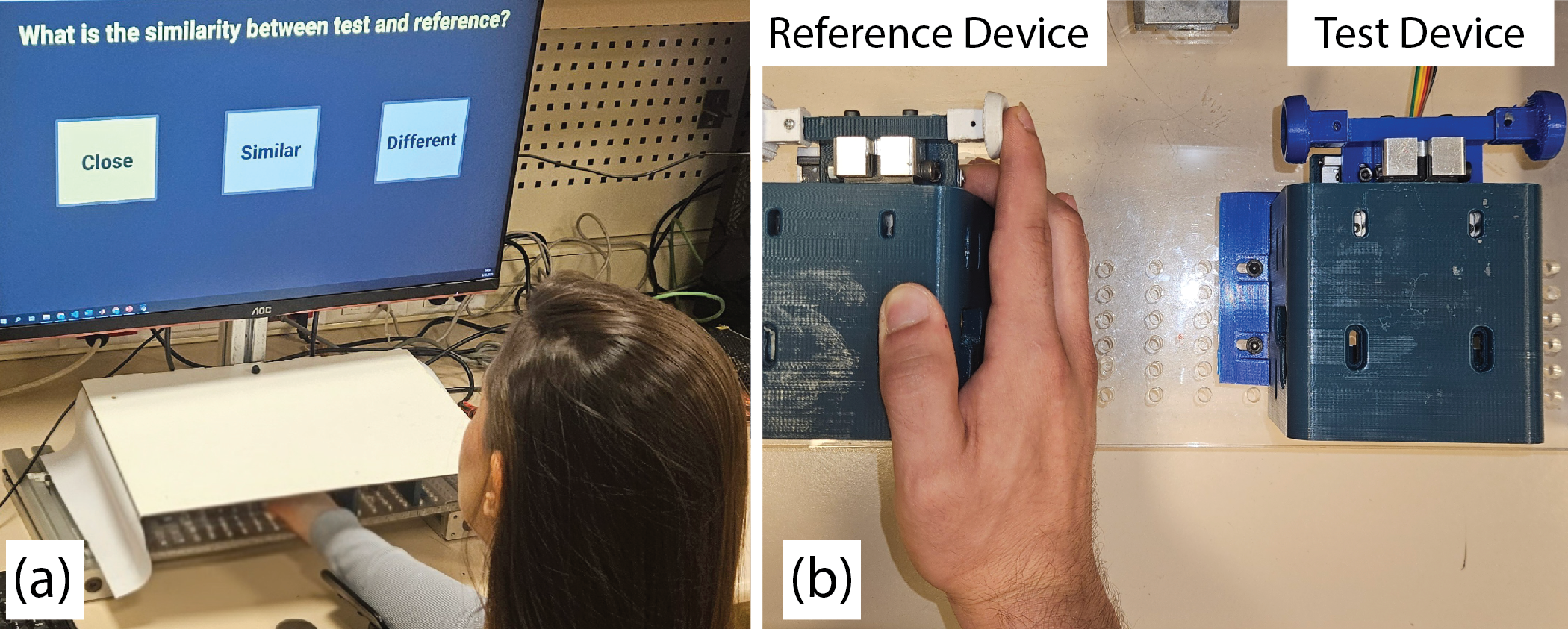}
    \vspace{-2mm}
    \caption{(a)~The experimental setup with a participant. (b)~Two identical haptic interfaces: The reference device with the physical viscoelastic material is on the left, while the test device rendering a viscoelastic model is on the right.}
    \label{fig:Haptic Device}
\end{figure}

The experimental setup consisted of two identical haptic interfaces and an interactive graphical user interface (GUI), as shown in Figure~\ref{fig:Haptic Device}. Although mechanically identical, the two devices had different roles during the experiments. One device served as the \emph{test device} for rendering the fractional-order viscoelastic models. The other device served as the \emph{reference device} and contained the \emph{physical} viscoelastic material as ground truth. As shown in Figure~\ref{fig:Haptic Device}(b), participants positioned their dominant hands sideways to hold the haptic devices with their palms facing inward and used their index fingers during interactions. This motion simulates the palpation motion commonly used in diagnostic procedures~\cite {palpation}.

Each haptic interface in Figure~\ref{fig:Haptic Device}(b) consists of a BEI-Kimco LA28-15-002Z linear actuator attached to a US Digital transmissive optical encoder with a 2000 counts/inch resolution. The device's frame was 3D-printed to facilitate comfortable handling and operation. High-precision linear guides were incorporated to minimize friction during operation.

For haptic rendering with the \emph{test device}, open-loop impedance control was implemented in real-time at 1~kHz using the \textsc{MATLAB} Real-Time environment.

\vspace{-2mm}
\subsection{Experimental Protocol}

\smallskip
\subsubsection{Setup and Overview}

All participants completed the experiment using their dominant hands. Participants were free to explore the systems as they wished. To minimize the influence of auditory cues, they wore foam earplugs to block outside noise. To eliminate potential visual cues, both haptic interfaces were covered throughout the experiment, as shown in Figure~\ref{fig:Haptic Device}(a). Participants used an arm stand to ensure comfortable, stable arm movements during the tests. During all sessions, the device containing the physical viscoelastic material served as the reference for comparisons. 

\smallskip
\subsubsection{Procedure}
\label{sec:Procedure}

Participants were instructed to compare the similarity between the virtual viscoelastic models rendered by the controlled haptic interface and the reference haptic interface with the physical viscoelastic material. They provided feedback based on perceived realism to the reference.

Each optimization trial followed a standardized experimental procedure. In each trial, parameter sets were sampled according to the UCB acquisition function. At the beginning of each trial, participants interacted with the reference haptic interface, followed by interaction with the virtual environment rendered on the other haptic interface.  There was an 8~s time limit for each trial. After the interactions, participants evaluated the similarity between the two interfaces by responding to the following question: “What is the similarity between the test and the reference?” They selected one of the following:

\begin{itemize}
\item \emph{Close}: The tested haptic interface feels the same or highly close to the reference haptic interface.
\item \emph{Similar}: The tested haptic interface feels moderately similar to the reference haptic interface.
\item \emph{Different}: The tested haptic interface feels significantly different from the reference haptic interface.  
\end{itemize}

Responses were submitted through a GUI. The feedback collected from participants was then used to update the GP model for perceived realism relative to the reference. During the experiments, participants were allowed to take short breaks whenever they felt fatigued.

\vspace{-2mm}
\subsection{Sessions}

Three types of experimental sessions were conducted with preceding \emph{warm-up} periods: \textit{Viscoelastic Model Optimization}, \textit{Model Validation}, and \textit{Aggregate-Best Validation}. These sessions followed the same general structure across different viscoelastic materials, with material-specific validation details described in the Supplementary Document~\cite{Supplementary}.

\subsubsection*{Warm-Up}

The warm-up period was designed to familiarize participants with the haptic rendering task and the haptic interface. During this period, all the participants experienced at least 12 different haptic renderings. Additional renderings were provided on request until the participants felt comfortable with the task. The warm-ups lasted about 4~min.

\smallskip
\subsubsection{Viscoelastic Model Optimization}

This session aimed to optimize the parameters of the \textit{fractional-order SLS model} to enhance its perceived realism relative to the reference material. Each participant interacted with the reference device with physical viscoelastic material and the haptic interface rendering a fractional-order SLS model for 8~s. Next, qualitative feedback was provided on how similar or different the two sensations felt. 
This response was used to update the acquisition function within the Bayesian optimization framework, and the optimization recommended the next set of model parameters to test. The optimization process converged on the optimal model parameters for each participant after 25 iterations. These experiments were conducted separately for the three viscoelastic materials. Including the warm-up period, the session lasted about 15~min for each material.

\begin{figure*}[t!]
\centerline{\includegraphics[width=1\textwidth]{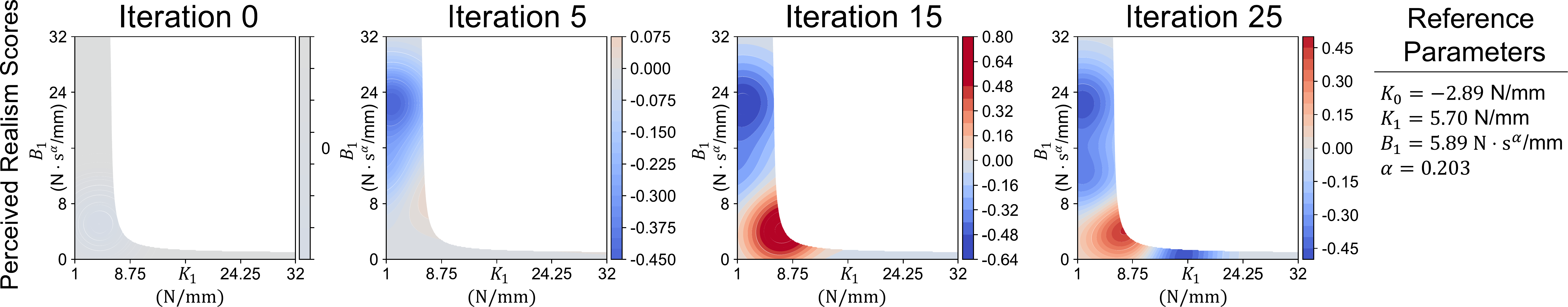}}
\vspace{-.1\baselineskip}
\caption{The progression of the posterior GP model of perceived realism depicted at various trials of the HiL optimization for a participant. For the presentation, the fractional order coefficient $\alpha$ is taken as a constant at $\alpha=0.19$, and two-dimensional slices in $K_1$ and $B_1$ axes are depicted. The white area represents a non-passive region omitted during sampling.} \vspace{-2mm}
\label{fig:iterations}
\end{figure*}

\smallskip
\subsubsection{Model Validation}

After training the GP models, \textit{Model Validation} sessions were conducted to evaluate whether the learned perceptual maps identified rendering parameters consistent with the human judgments. Three rendering parameter sets, labeled \emph{best}, \emph{mid}, and \emph{worst}, were selected from the posterior GP model to be validated. The \emph{best} and \emph{worst} sets corresponded to a local maximum and a local minimum of the posterior perceived-realism score, respectively, while the \emph{mid} set was selected to have an intermediate posterior score between them. For the memory-foam reference, these sets were selected from the aggregate GP model, as this session serves as the primary validation of the aggregate perceptual map. For the other silicone materials, an analogous post-optimization validation was conducted using parameter sets selected from each participant's individually trained GP model, whose details are presented in the Supplementary Document~\cite{Supplementary}.

During each validation trial, participants interacted with two selected renderings for 12~s and compared them in a forced-choice task. Participants classified the perceived similarity of each rendering to the physical reference as \emph{Close}, \emph{Similar}, or \emph{Different}, and then selected the rendering that felt more similar to the reference. The validation dataset contained preferences and classifications collected over 12~trials per participant, with each rendering type presented 8~times. The validation session, including the warm-up period, lasted approximately 10~min for each material.

\smallskip
\subsubsection{Aggregate-Best Validation}

In this session, the population-level aggregate-best parameter sets were compared against individually HiL-optimized best parameter sets in a forced-choice task across all three materials. In each trial, participants interacted with two models for 12~s and were asked to prefer the parameter set that felt more similar to the physical reference. Aggregate-Best Validation for the memory-foam included 24~individual-best parameter sets repeated twice, yielding 48~trials, while the validation for the silicone materials included 8~individual-best parameter sets repeated three times, yielding 24~trials per viscoelastic material. Participants completed validations in a randomized order, with the presentation of the memory foam and silicone materials counterbalanced to reduce ordering effects. Participants completed the warm-up once and took 10~min breaks while switching between different material types.  Aggregate-Best Validation for the memory foam and the silicone materials each lasted around 25~min.

\vspace{-2mm}
\subsection{Hypotheses} \label{Sec:hypothesis}

The following hypotheses were tested: \smallskip

\begin{itemize}
 \item {H1:} The parameters of the fractional-order SLS model can be effectively identified using a HiL framework utilizing Bayesian optimization with qualitative feedback for individuals, such that the rendering with these customized parameters has high perceived realism. \smallskip

 \item {H2:} HiL optimized perceptual mappings of a group of individuals can be effectively combined to determine a (population-level) aggregate perceptual map trained on the entire dataset,  and the population-level optimal parameters selected from this aggregate map are broadly perceived as realistic across general populations. \smallskip

 \item {H3:} The perceptual mappings (GP models) of individuals need not converge for the aggregate perceptual map trained on the entire dataset to effectively determine population-level optimal parameters that are perceived as realistic across general populations.  

 \item {H4:} The proposed HiL optimization and aggregation framework generalizes across viscoelastic reference materials with different mechanical properties.

\end{itemize}

\begin{figure}[b!]
    \centering
\begin{tabular}{cc}
\includegraphics[width=.23\textwidth]{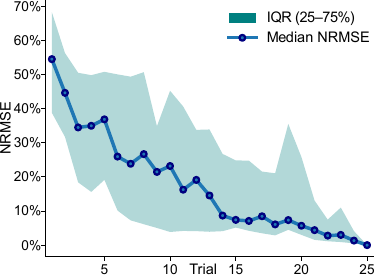} & 
\includegraphics[width=.23\textwidth]{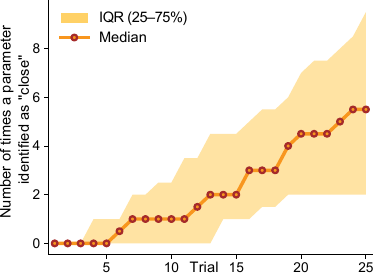} \\    
\end{tabular}
    \vspace{-1.5mm}
    \caption{(a)~NRMSE between each participant’s current best predicted parameter set versus their final best predicted parameter set; (b)~Number of times a participant evaluated a parameter set as \emph{close}, up to the current trial.}
    \label{fig:RMSE}
\end{figure}

\section{Results}

For brevity, the results of \emph{Viscoelastic Model Optimization}, \emph{Model Validation}, and \emph{Aggregate-Best Validation} experiments are presented only for the memory foam material, while similar results for the medium and high stiffness silicones are available in the Supplementary Document~\cite{Supplementary}.

For each individual, the active learning framework constructed a GP latent model of the perceived realism of the rendering parameters with respect to the reference. The progression of the mean of the GP model of a sample participant over the HiL optimization experiment trials is presented in Figure~\ref{fig:iterations}. Using the latent model, one can assign a perceptual realism score to any parameter set within the feasible design space employing Gaussian regression, even if that particular parameter set has not been directly tested.

\begin{figure*}[t!]
\centerline{\includegraphics[width=1\textwidth]{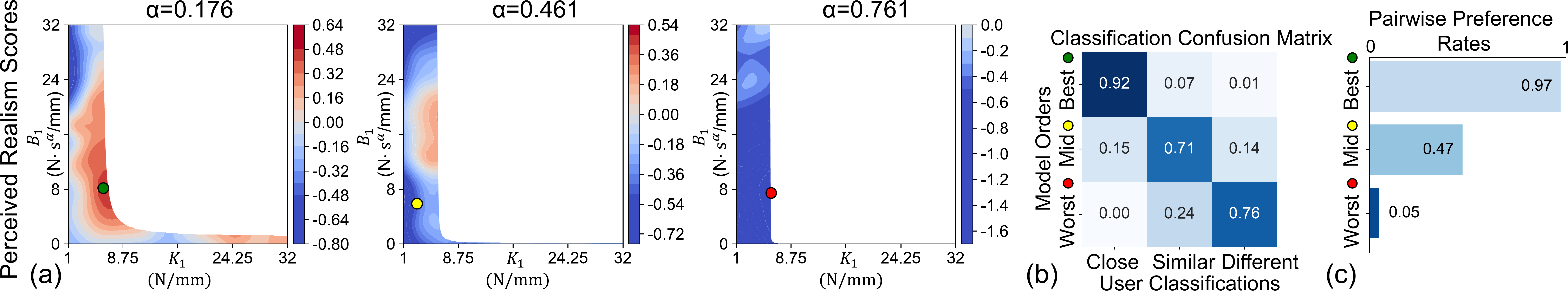}} \vspace{-1mm}
\caption{Different slices from the posterior of the aggregate GP model with fixed $\alpha$ levels. The green, yellow, and red points mark equidistant perceived realism scores in the search space, corresponding to the local maximum (\emph{best}), intermediate (\emph{mid}), and local minimum (\emph{worst}) parameter sets. (b)~Confusion matrix of user classifications across the three parameter sets. (c)~Pairwise preference rates for selecting the rendering that feels closer to the reference material, computed as the normalized number of times each parameter set is chosen over another in pairwise comparisons.}\vspace{-3mm} 
\label{fig:BCM_validation} 
\end{figure*}

Figure~\ref{fig:RMSE}(a) shows the evaluation of the error between the optimal parameter set at each trial with respect to the optimal values reached at the final trial, where NRMSE is computed using the normalized parameter values. 

Figure~\ref{fig:RMSE}(b) presents the number of times participants evaluated a parameter set as \emph{close}, up to the current trial. Out of 25 trials, the participants on average evaluated $5.96\pm4.47$ trials as \emph{close}. In Figures~\ref{fig:RMSE}(a) and~\ref{fig:RMSE}(b), the interquartile range~(IQR) indicates the variability across the participants.

Figure~\ref{fig:BCM_validation}(a) presents different slices of the posterior of the aggregate GP model. The green, yellow, and red points mark the equidistant perceived realism scores in the search space, corresponding to the local maximum (called \emph{best}), local minimum (called \emph{worst}), and intermediate (called \emph{mid}) rendering parameter sets of the aggregate GP posterior model.

The aggregate GP posterior model estimates the population-level optimal parameter set, marked with a green dot in Figure~\ref{fig:BCM_validation}(a), as given in Table~\ref{tab:aggregate_vs_sysid}. These parameters maximize the perceived realism score under the aggregate GP posterior model and closely match those obtained from system identification for the memory foam material. The errors in Table~\ref{tab:aggregate_vs_sysid} are normalized with respect to the search-space.

In contrast, the statistical average of individual best parameters can be computed as\(K_0 = -5.01 \pm 3.82~\text{N/mm},\ K_1 = 8.44 \pm 8.25~\text{N/mm},\ B_1 = 10.28 \pm 8.64~\text{N·s}^{\alpha}\!/\text{mm},\ \alpha = 0.282 \pm 0.140\). These averages exhibit larger normalized discrepancies with respect to the system identification parameters, with normalized percentage errors of \(13.1\%\) for \(K_0\), \(8.8\%\) for \(K_1\), \(13.7\%\) for \(B_1\),  and \(8.1\%\) for \(\alpha\), respectively.

\begin{table}[t]
\centering
\caption{System identification vs. aggregate GP parameters} \vspace{-2mm}
\label{tab:aggregate_vs_sysid}
\small
\begin{tabular}{l|c|c|c} 
\hline
Parameter & Sys. ID & Agg. GP & Error (\%) \\
\hline
$K_0$~(N/mm)
& $-2.89$ & $-3.18$ & $1.8$ \\
$K_1$~(N/mm)
& $5.70$  & $5.58$  & $0.4$ \\
$B_1$~($\mathrm{N\,s^\alpha/mm}$)
& $5.89$  & $7.88$  & $6.2$ \\
$\alpha$
& $0.203$ & $0.176$ & $2.8$ \\
\end{tabular} \normalsize \vspace{-2mm}
\end{table}

User responses collected during the Model Validation session for the memory foam, summarized in Figure~\ref{fig:BCM_validation}(b) and Figure~\ref{fig:BCM_validation}(c), indicate a clear and consistent ordering. The \emph{best} set is preferred over the \emph{mid} and \emph{worst} sets in 95\% and 100\% of comparisons, respectively, while the \emph{mid} set is preferred over the \emph{worst} set in 90\% of comparisons.

\begin{figure*}[t!]
\centerline{\includegraphics[width=0.8\textwidth]{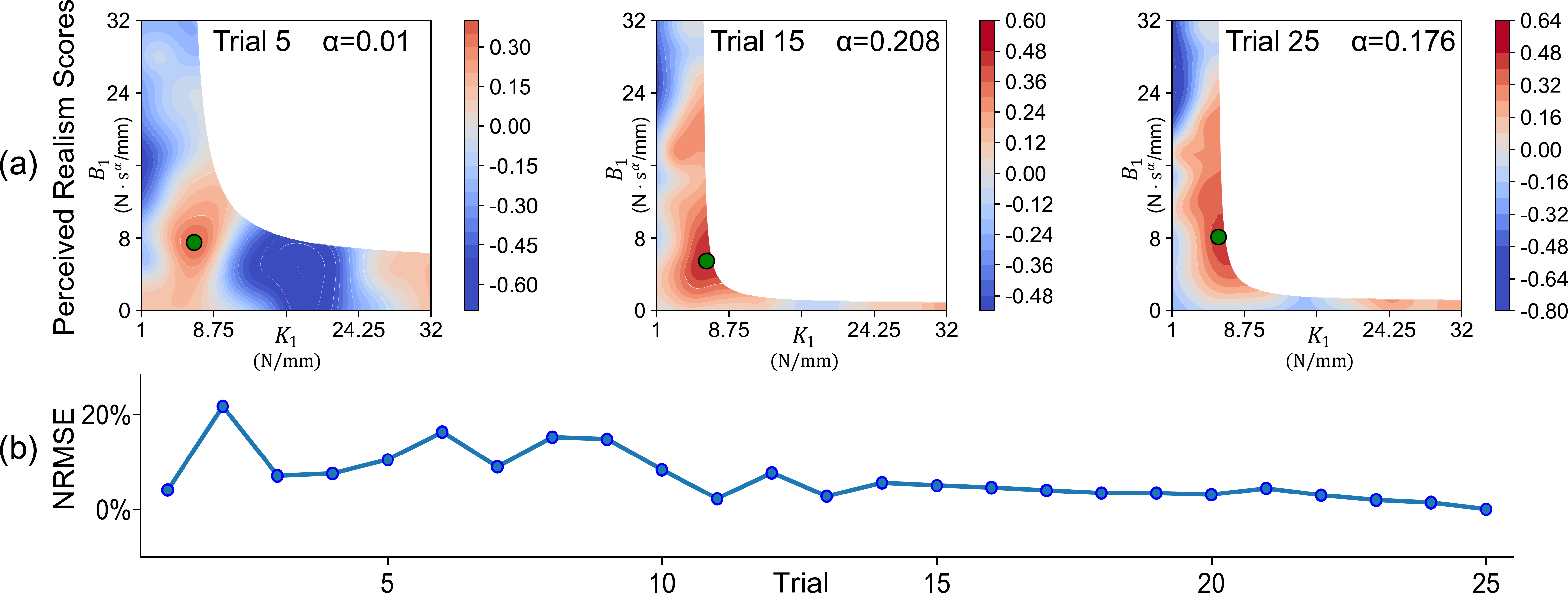}}\vspace{-1mm}
\caption{(a)~The estimates of the population-level optimal parameter sets, if the individual models were trained up to 5, 15, and 25 trials. (b)~The normalized prediction error between the optimal parameter sets estimated according to the aggregate GP models trained up to a given trial versus the optimal parameter sets estimated according to the final aggregate GP model.} 
\label{fig:Aggregate_Progress} \vspace{-2mm}
\end{figure*}

\begin{table}[t!]  \centering 
\caption{{Statistical testing for the classification preference of \emph{best} parameter set being \emph{close} or \emph{not close} to the reference}}\label{Table:3} \vspace{-2mm}
\begin{tabular}{l|c|c|c|c}
               & estimate & std. error & t-ratio & p-value     \\ \hline
Intercept & -2.5123   & 0.213     & -11.794   & $<\!0.001$\\ \hline
Best & 4.94   & 0.67     & 7.36   & $<\!0.001$
\end{tabular} \vspace{-1mm} 
\end{table}

\begin{table}[t!]\centering 
\caption{{Statistical testing for the difference between \emph{mid} and other parameters}}\label{Table:4} \vspace{-1.5mm}
\begin{tabular}{l|c|c|c|c}
      & estimate & std. error & t-ratio & p-value  \\ \hline
Best   & -4.1431   & 0.7104     & -5.832   & $<0.001$ \\ \hline
Worst & 2.9738   & 0.4034     & 7.371   & $<0.001$
\end{tabular}
\end{table}

\begin{table}[t!]\centering
\caption{Statistical testing for the preference of the aggregate-best rendering over individual-best renderings.}\label{Table:5} \vspace{-1.5mm}
\begin{tabular}{l|c|c|c|c}
      & estimate & std. error & t-ratio & p-value  \\ \hline
Aggregate-Best  & 0.9773   & 0.0844    & 11.5809   & $<0.001$
\end{tabular}
\end{table}

The classification data of \textit{Model Validation} session were first analyzed by recoding the ordinal responses into a binary outcome, with \emph{close} coded as~1 and \emph{not close} coded as~0, where the latter combined the \emph{similar} and \emph{different} responses. A binary logit model was fitted to test whether the \emph{best} parameter set was more likely to be classified as \emph{close} to the reference. The results in Table~\ref{Table:3} indicate that the \emph{best} parameter set was significantly more likely to be classified as \emph{close}.

To further analyze the ordinal classification judgments, an ordered logistic regression model was fitted using the original ordered responses, coded as \emph{close} = 1, \emph{similar} = 2, and \emph{different} = 3. The \emph{mid} parameter set was used as the reference level; therefore, a negative coefficient indicates a shift toward the \emph{close} category, whereas a positive coefficient indicates a shift toward the \emph{different} category. The results in Table~\ref{Table:4} show that the \emph{best} and \emph{worst} parameter sets were significantly differentiated from the \emph{mid} set in the expected directions.

The \textit{Aggregate-Best Validation} session tested whether the population-level aggregate-best parameter set remained favorable when directly compared with individually optimized best parameter sets. For the memory-foam reference, the aggregate-best rendering was preferred in 72.7\% of the pairwise comparisons against individually optimized best renderings.

The pairwise preference data from the \textit{Aggregate-Best Validation} session were analyzed using another binary logit model, in which preferences for the aggregate-best and individual-bests were coded as~1 and~0, respectively. The results in Table~\ref{Table:5} indicate that the aggregate-best rendering was preferred significantly more often than the individually optimized~ones.

In all statistical analyses, cluster-robust standard errors were used to account for repeated responses within participants; for the aggregate-best analysis, two-way cluster-robust standard errors were used to account for repeated evaluations within participants and repeated use of the same individual-best parameter sets across participants.

To complement the statistical analysis, we also quantified how closely the estimated best parameters matched the parameters obtained from system identification. For this purpose, the normalized distances between the parameters determined via system identification and the best-predicted parameters of the individual and aggregate GP models were computed. The error between the best estimate of the aggregate GP model and normalized values of the system identification parameters resulted in $4\%$~NRMSE, while the error between best predicted estimates of individually trained GP models and the normalized parameters due to system identification has an IQR between $13\%$ and $41\%$ with a median of $24\%$. 

Finally, Figure~\ref{fig:Aggregate_Progress} presents the effect of the number of trials during the experiment on the convergence of the aggregate GP model. Figure~\ref{fig:Aggregate_Progress}(a) presents the mean of the aggregate GP models if the individual models were trained up to 5, 15, and 25 trials. The locations of the estimated best parameter sets with the highest perceived realism scores in the corresponding trials are depicted as green circles, with slices taken at the $\alpha$ levels of the corresponding optimal parameter set. Figure~\ref{fig:Aggregate_Progress}(b) shows the prediction error between the optimal parameter sets estimated according to the aggregate GP models trained up to a given trial versus the optimal parameter sets estimated according to the final aggregate GP model.

To further analyze the aggregate perceptual map, we consider the posterior probability of ordinal classifications for a given rendering parameter set, defined as \vspace{-2mm}

{\small
\begin{equation*}\label{eq:Aggregate Class Prob}
\!P(q_*=o_j|\boldsymbol{D_{tot}},x_*) \!=\! \Phi\!\left(\!\frac{t_j-\mu_{*|\boldsymbol{D_{tot}}}}{\sqrt{c_o^2+\sigma^2_{*|\boldsymbol{D_{tot}}}}}\!\right)\!-\Phi\!\left(\!\frac{t_{j-1}-\mu_{*|\boldsymbol{D_{tot}}}}{\sqrt{c_o^2+\sigma^2_{*|\boldsymbol{D_{tot}}}}}\! \right).
\end{equation*}}
\normalsize

\noindent Here, $q_*$ denotes the ordinal decision corresponding to an arbitrary rendering parameter set $x_*$, and $o_j$ denotes the ordinal class, where $o_3$ corresponds to the \emph{close} category. Based on this formulation, we define a \emph{close}-dominant region for the set of parameters predicted more likely to be classified as \emph{close} rather than \emph{similar} or \emph{different}.

\begin{figure}[t!]
    \centering
    \includegraphics[width=.9\linewidth]{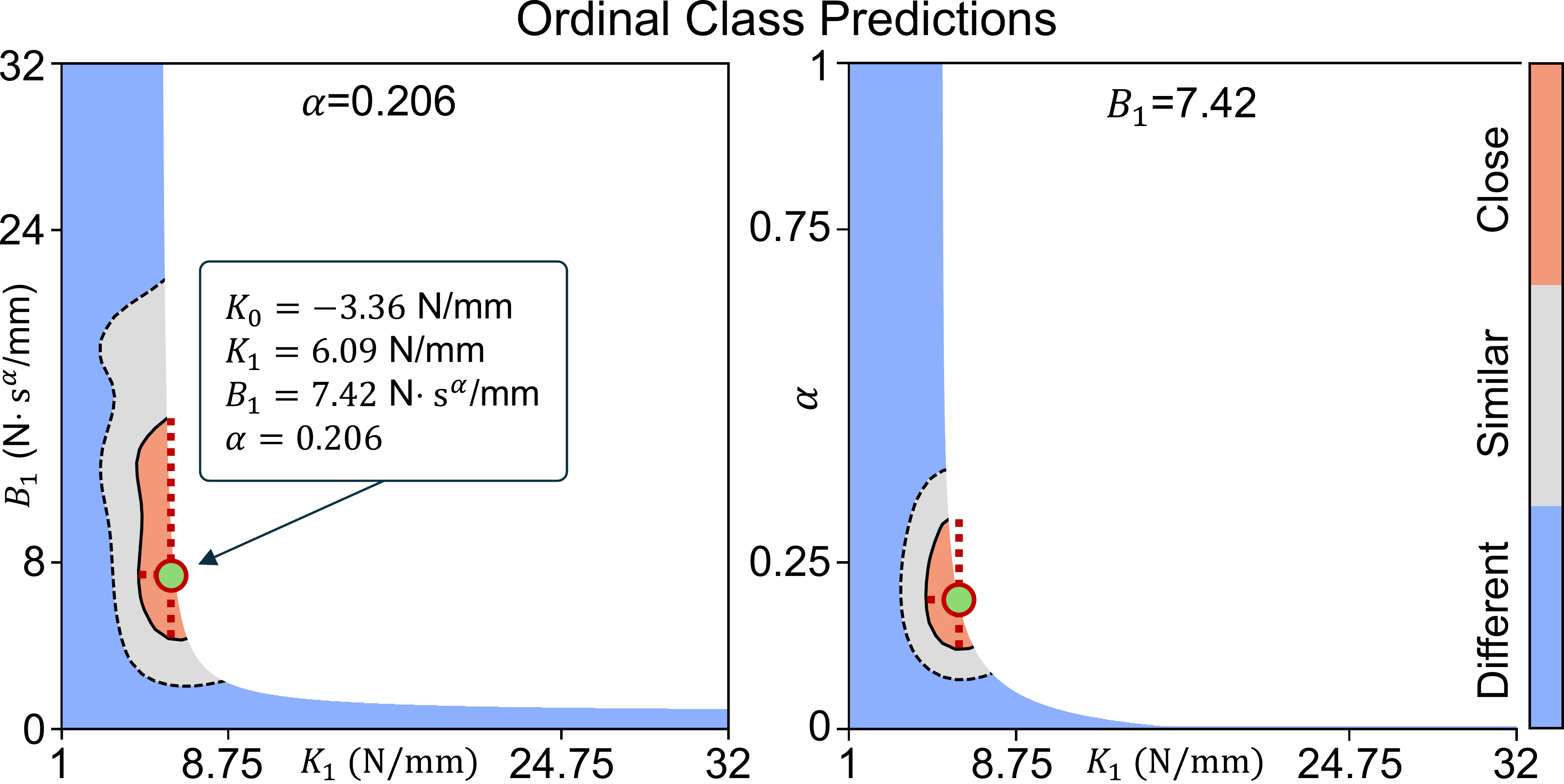} \vspace{-1mm}
    \caption{Ordinal class predictions of the aggregate GP model. $(K_1, B_1)$ and $(K_1, \alpha)$ slices intersect at the best predicted parameter set, shown by the green marker. Regions denote the most probable ordinal class, and red dashed arrows indicate the directions used for the perceptual sensitivity analysis.}
    \label{fig:Ordinal Classification Mapping}
\end{figure}

To examine the local structure of the \emph{close}-dominant region, we analyzed one-dimensional slices of the aggregate GP model with optimized hyperparameters, as shown in Figure~\ref{fig:Ordinal Classification Mapping}. Along each slice, one parameter was varied while the others were fixed near the population-level best-predicted rendering parameter set. The interval over which the \emph{close} category remained the most probable ordinal class was then identified. Since the aggregate optimum lies near the passivity boundary, these intervals were computed only over feasible passive regions: for $B_1$ and $\alpha$, the closest passive slices to the aggregate optimum were used, while for $K_1$, a one-sided interval toward lower stiffness values was reported because higher $K_1$ values violate passivity. The resulting intervals were $K_1 \in [4.57,\,6.09]$ N/mm, $B_1 \in [4.33,\,15.09]$ N$\cdot$s$^\alpha$/mm, and $\alpha \in [0.128,\,0.327]$.

\vspace{-1mm}
\section{Discussion}
In this section, we elaborate on the validity of each hypothesis presented in Section~\ref{Sec:hypothesis} and discuss the sensitivity perception to parameter changes.

\subsection{Hypothesis~1}

As illustrated in Figures~\ref{fig:RMSE}(a) and~(b), NRMSE between the current best predicted parameter and the final best predicted parameter decreases significantly during the first 15~trials, while the median number of parameter sets evaluated as \emph{close} to the reference increases across trials. These complementary trends indicate that the algorithm progressively identified and refined a promising region of the parameter space with high perceived realism. Additionally, IQRs start to converge after about 20~trials, suggesting increasing consistency across participants as they converged towards regions with high perceived realism. The median trend in Figures~\ref{fig:RMSE}(a) and~\ref{fig:RMSE}(b) also indicates that the last 5~trials were dominantly used to fine-tune parameters within an already identified promising region. This behavior reflects the expected exploration-exploitation behavior of the HiL optimization. Overall, these trends suggest that the GP models of most participants converged to an optimal or near-optimal region within 25~trials, consistent with earlier HiL Bayesian optimization studies~\cite{Catkin2023, Tolasa2024}, and that 25~trials provided an appropriate balance between searching the parameter space and inducing fatigue in participants.

While the median plots indicate an overall convergence to parameters with high perceived realism, IQRs in Figures~\ref {fig:RMSE}(a) and~\ref{fig:RMSE}(b) indicate that GP models of some participants did not fully converge to parameters set with high perceived realism within 25 trials. Such convergence behavior is expected for two reasons: First, perceptual feedback naturally involves variability; participants may become momentarily uncertain about their decisions, or their internal evaluation strategy may shift over the course of a session. Such inconsistencies can cause the GP model to shift to or plateau at a non-optimal region. Second, unlike prior HiL studies operating in one- or two-dimensional search spaces, the current task requires exploration in a three-dimensional parameter space, making convergence naturally more challenging in higher-dimensional spaces. Overall, achieving robust convergence for all participants under subject variability and large search spaces requires more trials in the HIL context, which would cause (mental or physical) fatigue for participants.

Luckily, convergence of GP models across all participants is \emph{not} required to obtain a reliable population-level estimate. This aspect is further discussed in detail in Section~\ref{H3}.

Overall, the results in Figures~\ref{fig:RMSE}(a) and~\ref{fig:RMSE}(b) strongly support the hypothesis~H1; for the majority of the participants, the active learning framework successfully identified fractional-order SLS model parameters with high perceived realism relative to the reference.

\subsection{Hypothesis~2} 

The close correspondence between the population-level optimal parameters determined by the aggregate GP model and those from system identification provides strong evidence that HiL-optimized perceptual mappings from a group of individuals can be combined into an aggregate perceptual map that captures realistic interaction dynamics across general populations.

For the aggregate-best parameters, the normalized errors remained small for all parameters: 1.8\% for $K_0$, 0.4\% for $K_1$, 6.2\% for $B_1$, and 2.8\% for $\alpha$. These errors were considerably smaller than those obtained from the statistical average of individual best parameters, which yielded errors of 13.1\%, 8.8\%, 13.7\%, and 8.1\%, respectively. Similarly, the aggregate-best parameters yielded a 4\% NRMSE relative to the system-identification parameters, whereas the individually optimized parameters exhibited substantially higher variability, with an IQR of 13--41\% and a median of 24\%. These quantitative comparisons indicate that the BCM-based aggregate GP model captures the population-level perceptual consensus more reliably than averaging individual best parameter estimates.

In the literature, statistical averaging of posterior latent functions has been used to construct perceptual maps for broader populations and estimate model parameters with the highest perceived realism~\cite{Tolasa2024}. When the perceptual space is well sampled, and individual GPs provide low-uncertainty predictions across comparable regions of the domain, statistical latent-function averaging and BCM-based aggregation are expected to yield similar estimates. However, for higher-dimensional search spaces with limited trial budgets, individual GP models often contain sparsely or inconsistently sampled regions with high predictive variance. Treating these uncertain regions the same as well-sampled ones can dilute or distort the population-level estimate. Consequently, a rigorous aggregation method, such as BCM, is advantageous because it weights individual predictions by their precision.

The Model Validation results indicate that the aggregate GP model (perceptual map) trained on the entire dataset successfully captures perceived realism, as evidenced by the diagonal dominance in the confusion matrix in Figure~\ref{fig:BCM_validation}(b). The pairwise preferences in Figure~\ref{fig:BCM_validation}(c) further show that the aggregate model adequately ordered the \emph{best}, \emph{mid}, and \emph{worst} parameter sets according to human perception.

Table~\ref{Table:3} indicates that the regression produces a strong estimate for both the intercept $(\beta_{inter} = -2.51\pm 0.21)$ and the best parameter set $(\beta_{best} = 4.94\pm 0.67)$. Moreover, the estimate of the \emph{best} parameter set implies a strong correlation for evaluating the \emph{best} parameter set as \emph{close} to the reference, while the estimate of the intercept implies that there is a strong correlation in selecting other parameter sets as \emph{not close}. Consequently, the null hypothesis, stating that the \emph{best} parameter set does not affect user decisions while classifying the perception as \emph{close} or \emph{not close} to the reference, can be rejected with statistical significance at the $p<0.001$ level.

Table~\ref{Table:4} shows that when the \emph{mid} parameter set was used as the reference level, both the \emph{best} and \emph{worst} parameter sets showed statistically significant differences at the $p < 0.01$ level. The estimates of the \emph{best} parameter set being negative ($\beta_{best} = -4.14 \pm 0.71$) and the \emph{worst} parameter set being positive ($\beta_{worst} = 2.97 \pm 0.40$) confirm that all parameter sets were distinguishable and ordered consistently with their underlying orderings, whereas the \emph{best} parameter set was perceived as the closest to the reference among the three~sets.

The \textit{Aggregate-Best Validation} further helps evaluate the validity of the hypothesis~H2 by testing the practicality of the aggregation by directly comparing the aggregate-best parameter set against individually optimized parameter sets. For the memory-foam material, the aggregate-best rendering was preferred in 72.7\% of the pairwise comparisons, and the binary logit analysis confirmed that this preference was statistically significant at $p<0.001$ level. These results indicate that the aggregate perceptual map identifies a population-level solution rather than simply averaging over user-specific optima.

Overall, both the \textit{Model Validation} and the \textit{Aggregate-Best Validation} results for the memory-foam material, together with the statistical analyses in Tables~II--IV, provide strong evidence that the hypothesis~H2 holds, indicating that HiL-optimized GP models from a group of individuals can be effectively combined to determine an aggregate perceptual map and this map can be used to determine a population-level optimal parameter set that is broadly perceived as realistic.

\vspace{-1mm}
\subsection{Hypothesis~3} \label{H3}

Figure~\ref{fig:Aggregate_Progress} illustrates how the aggregate GP model converges as the number of trials is increased. In Figure~\ref{fig:Aggregate_Progress}(a), the aggregate GP models progressively stabilize, and the location of the predicted optimal parameter set remains within a small region after the early trials. Figure~\ref{fig:Aggregate_Progress}(b) further quantifies this effect by computing NRMSE between the optimal parameter sets estimated from intermediate aggregate GP models and the final aggregate GP model. NRMSE decreases rapidly in the first 11 trials and then approaches a plateau, indicating that additional trials primarily refine the aggregate model without substantially altering the predicted population-level optimum.

More importantly, results in Figure~\ref{fig:Aggregate_Progress} provide strong evidence that the aggregate GP model trained on the entire dataset can converge even when individual participants' GP models do not converge. In particular, high standard deviations of the statistically averaged model reported in Figure~\ref{fig:RMSE}(a) indicate that individual GP models of some participants did not fully converge, while Figure~\ref{fig:Aggregate_Progress} shows definite convergence of the aggregate GP model. Such a result is expected because, as long as the collective samples from all participants sufficiently cover the search space, the aggregate GP model constructed through the BCM framework will reliably converge to a stable, representative population-level optimum by integrating consistent perceptual trends across participants. 

Accordingly, if the goal is to determine population-level optimal parameters that are perceived as realistic across general populations, then HiL data collected from many individuals can be aggregated without requiring convergence of individual HiL optimizations. By exploiting shared information across partially converged individual GPs using BCM, the aggregate GP model can efficiently capture the perceptual trends without requiring every individual model to fully converge. In practice, to provide the information needed for BCM to assess the population-level prediction, it is sufficient for at least some participants to reach and correctly evaluate the optimal region. This aspect of our proposed approach is crucial for the feasibility of obtaining realistic parameters for models with a large number of parameters, for which imposing convergence of individual HiL optimizations becomes infeasible due to the large size of the search space, which can cause participants to become mentally or physically fatigued before full convergence can be achieved. 

Overall, our observations confirm that the aggregate GP model can achieve reliable convergence even when all individual participant models have not yet fully converged. These results support our hypothesis~H3 that the perceptual mappings of individuals do not necessarily need to have converged in order for the aggregate GP model to effectively represent a generalized and realistic perception across the population.

\vspace{-1mm}
\subsection{Hypothesis~4} \label{H4}

The additional experiments with the medium- and high-stiffness silicone materials provide evidence of the generalizability of the proposed framework across viscoelastic materials with different mechanical properties. The same material characterization and HiL optimization procedures were applied to these additional silicone materials, and the results reported in the Supplementary Document~\cite{Supplementary} show that the individually trained GP models produced perceptually meaningful \emph{best}, \emph{mid}, and \emph{worst} rendering conditions. In addition, the \textit{Aggregate-Best Validation} results show that the population-level aggregate-best renderings remained favorable when directly compared with individually optimized renderings. These findings support the hypothesis~H4 by indicating that the proposed approach is not limited to the original memory-foam reference and can be applied to viscoelastic materials with different mechanical characteristics, provided they can be faithfully represented by the fractional-order SLS model.

While these results provide strong evidence for the hypothesis~H4, the performance of the framework also depends on the consistency of qualitative feedback. If the relevant viscoelastic cues are close to the perceptual sensitivity limits, one cue dominates the overall judgment,  multiple cues contribute ambiguously to perceived realism, user responses may become noisier, reducing the reliability of the learned perceptual map. Therefore, the framework is expected to perform reliably when the feasible parameter range is selected to ensure users can provide consistent perceptual judgments.

\vspace{-2mm}
\subsection{Analysis of Perceptual Sensitivity}

Figure~\ref{fig:Ordinal Classification Mapping} provides an interpretable view of the aggregate GP model by converting the latent perceived-realism function into predicted ordinal response regions. Around the aggregate optimum, the \emph{close}-dominant region indicates the range of rendering parameters that are expected to preserve a high perceived similarity to the physical reference. The corresponding sensitivity intervals suggest that the population-level realism judgment is locally tolerant of moderate changes in the $B_1$ and $\alpha$ parameters, whereas variations in $K_1$ are more restricted because the passivity boundary constrains feasible values of this parameter. Accordingly, the aggregate model not only identifies a single best parameter set but also provides a local region of acceptable renderings to guide designers in selecting robust model parameters.

At the same time, these intervals should be interpreted as sensitivities specific to the underlying model and its parameters, rather than as a complete decomposition of viscoelastic perception into independent perceptual attributes. The parameters of the fractional-order SLS model jointly influence stiffness, damping, and time-dependent behavior, and different combinations of parameters may yield similar perceptual responses. A more detailed separation of the perceptual contributions of stiffness, damping, creep, and stress relaxation would require dedicated psychophysical experiments, as in~\cite{Tolasa2024}, in which specific feedback is collected separately for each attribute.

\vspace{-2mm}
\section{Conclusion and Future Work}

We proposed a systematic active learning approach to determine the parameters of fractional-order viscoelastic models for realistic haptic rendering of viscoelastic materials for the general population. We demonstrated that the viscoelastic behavior can be accurately modeled using fractional-order models, with parameters determined from qualitative user feedback. 

By aggregating the independently optimized perceptual maps of several participants, we developed a generalized fractional-order model that maintains a high level of perceived realism across participants. Validations through human-subject experiments provided strong evidence for the validity and robustness of our population-level optimal parameters in simulating realistic viscoelastic sensations.

Our findings indicate that the proposed HiL optimization and aggregation technique provides a systematic and practical solution for fractional-order viscoelastic model parameterization. The proposed approach possesses the potential to significantly enhance the perceived realism of medical and haptic training simulators.

This study demonstrates that the current framework can optimize three parameters within a feasible number of HiL trials. On the other hand, further efficiency improvements may be needed for higher-dimensional optimization tasks, as the search space grows exponentially with respect to the number of parameters. To this end, the efficiency of the framework could be extended by utilizing more informative sampling strategies~\cite{LineCospar2020}, richer qualitative feedback per trial~\cite{Li2020, Tolasa2024}, and warm starts from previously collected population-level perceptual data~\cite{Liao2025}.

Furthermore, while the present work demonstrates the capability of fractional-order models to describe realistic viscoelastic behavior, a systematic evaluation of perceived realism across models of increasing complexity remains to be conducted. We plan to compare models with varying numbers of parameters to assess their optimization costs and perceptual performance. Such an analysis will enable a better understanding of the trade-offs among computational efficiency, model complexity, and perceived realism in viscoelastic rendering.

\vspace{-3mm}
\section*{Acknowledgment}

This work has been partially supported by the TUBITAK Grant~23AG003. 

\vspace{-3mm}

\bibliographystyle{IEEEtran}
\bibliography{Bibliography}

\vspace{-1\baselineskip}
\begin{IEEEbiography}[{\includegraphics[width=.8in,height=1.2in,clip,keepaspectratio]{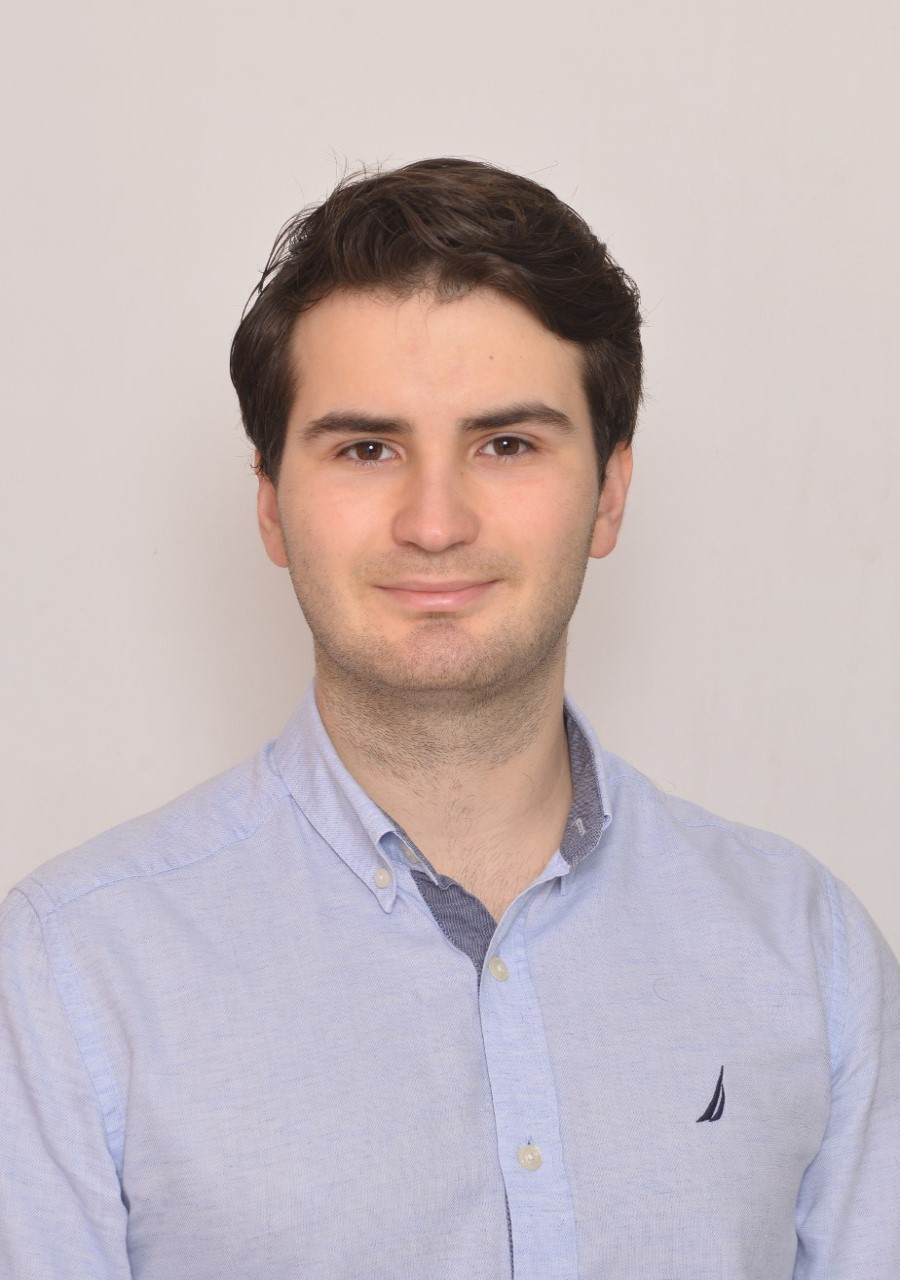}}]
{Harun Tolasa}  received his B.Sc. degree in mechanical engineering from Bilkent University (2021) and his M.Sc. in mechatronics engineering from Sabanci University~(2024). Currently, he is pursuing his Ph.D. degree at Sabanci University. His research interests include active learning, human-in-the-loop optimization, and haptic rendering.
\end{IEEEbiography}

\vspace{-1\baselineskip}
\begin{IEEEbiography}[{\includegraphics[width=.8in,height=1.2in,clip,keepaspectratio]{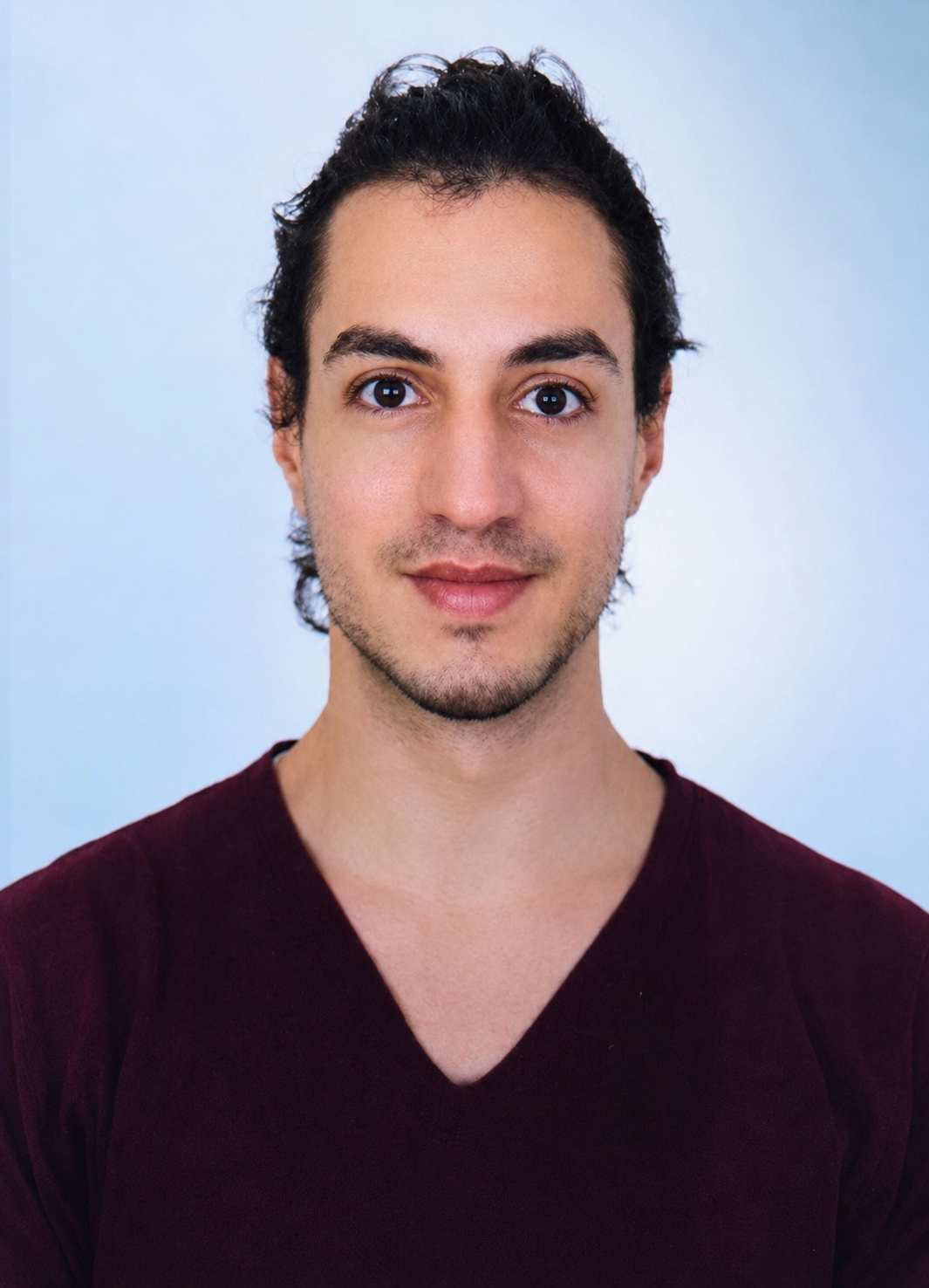}}]
{Gorkem Gemalmaz}  received his B.Sc. degree in mechanical engineering from Middle East Technical University~(2022). Currently, he is pursuing his Ph.D. degree at Sabanci University.
His research interests include physical human-robot interaction and interaction control.
\end{IEEEbiography}

\vspace{-1\baselineskip}
\begin{IEEEbiography}[{\includegraphics[width=1in,height=1.2in,clip,keepaspectratio]{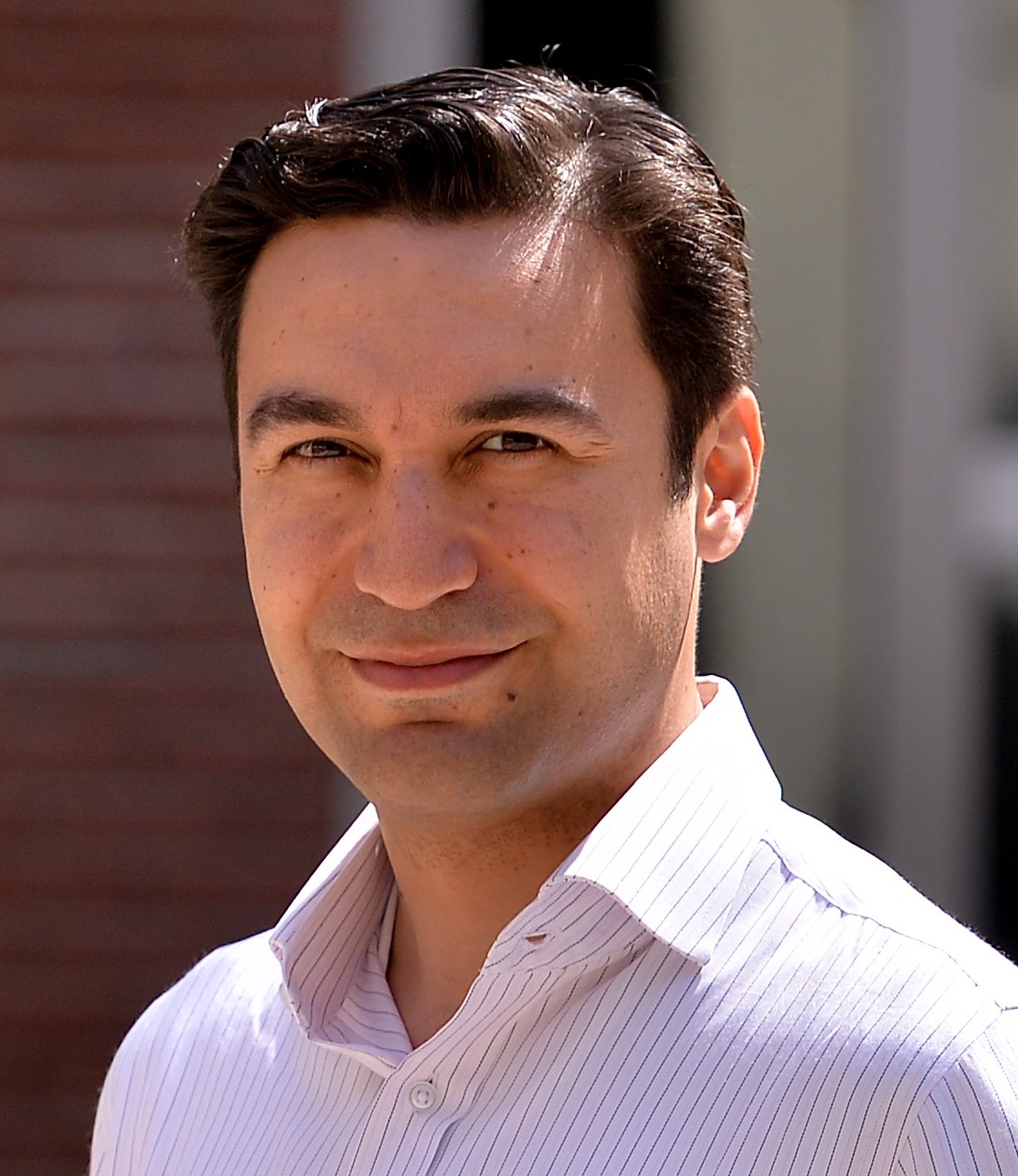}}]
{Volkan Patoglu} is a professor in mechatronics engineering at Sabanci University.
He received his Ph.D. in mechanical engineering from the University of Michigan, Ann Arbor~(2005) and worked as a postdoctoral researcher at Rice University~(2006). His research is in the area of physical human-machine interaction, in particular, the design and control of force-feedback robotic systems with applications in rehabilitation. His research extends to cognitive robotics. He has served as associate editor for IEEE Transactions on Haptics~(2013--2017), IEEE Transactions on Neural Systems and Rehabilitation Engineering~(2018--2023), and IEEE Robotics and Automation Letters~(2019--2024).
\end{IEEEbiography}

\vfill

\end{document}